\documentclass[]{spie}
\usepackage{comment}

\usepackage{amsmath,amsfonts,amssymb}
\usepackage{graphicx}
\usepackage[colorlinks=true, allcolors=blue]{hyperref}
\usepackage{comment}
\usepackage{schemabloc}
\usepackage{mathtools}
\usepackage{subcaption}
\usepackage{float}
\usepackage{makeidx}
\makeindex

\usepackage[pscoord]{eso-pic}

\title{The 2023 Balloon Flight of the ComPair Instrument}
\author[1,2]{Lucas D. Smith}
\author[2,3,4]{Nicholas Cannady}
\author[2]{Regina Caputo}
\author[2]{Carolyn Kierans}
\author[2,5]{Nicholas Kirschner}
\author[2]{Iker Liceaga-Indart}
\author[2]{Julie McEnery}   
\author[1,2,4]{Zachary Metzler}
\author[1,2,4]{A. A. Moiseev}
\author[6]{Lucas Parker}
\author[2]{Jeremy Perkins}
\author[1,2,4]{Makoto Sasaki}
\author[2]{Adam J. Schoenwald}
\author[7]{Daniel Shy}
\author[2,3,4]{Janeth Valverde}
\author[2,4,8]{Sambid Wasti}
\author[7]{Richard Woolf}
\author[9]{Aleksey Bolotnikov}
\author[10]{Thomas J. Caligiure}
\author[10]{A. Wilder Crosier}
\author[9]{Jack Fried}
\author[2,3,4]{Priyarshini Ghosh}
\author[11]{Sean Griffin}
\author[7]{J. Eric Grove}  
\author[2]{Elizabeth Hays}
\author[12]{Emily Kong}
\author[2]{John Mitchell}   
\author[7]{Bernard Phlips}
\author[7]{Clio Sleator}
\author[2]{D.J. Thompson}
\author[7]{Eric Wulf}
\author[2,3,4]{Anna Zajczyk}

\affil[1]{University of Maryland, College Park, MD 20742, USA}
\affil[2]{NASA Goddard Space Flight Center, Greenbelt, MD 20771, USA}
\affil[3]{University of Maryland Baltimore County, Baltimore, MD21250, USA}
\affil[4]{Center for Research and Exploration in Space Science and Technology, Greenbelt, MD 20771, USA}
\affil[5]{George Washington University, Washington, DC 20052, USA}
\affil[6]{Los Alamos National Laboratory, Los Alamos, NM 87545, USA}
\affil[7]{U.S. Naval Research Laboratory, Washington, DC 20375, USA}
\affil[8]{Catholic University of America, Washington, DC 20064}
\affil[9]{Brookhaven National Laboratory, Upton, NY 11973, USA}
\affil[10]{Naval Research Enterprise Internship Program, resident at U.S. Naval Research Laboratory, Washington,
DC 20375, USA}
\affil[11]{Wisconsin IceCube Particle Astrophysics Center, Madison, WI 53703, USA}
\affil[12]{Technology Service Corporation, Arlington, VA 22202, USA}

\begin{document}
\maketitle

\begin{abstract}

The ComPair balloon instrument is a prototype gamma-ray telescope that aims to further develop technology for observing the gamma-ray sky in the MeV regime. ComPair combines four detector subsystems to enable parallel Compton scattering and pair-production detection, critical for observing in this energy range. This includes a 10 layer double-sided silicon strip detector tracker, a virtual Frisch grid low energy CZT calorimeter, a high energy CsI calorimeter, and a plastic scintillator anti-coincidence detector. The inaugural balloon flight successfully launched from the Columbia Scientific Balloon Facility site in Fort Sumner, New Mexico, in late August 2023, lasting approximately 6.5 hours in duration. In this proceeding, we discuss the development of the ComPair balloon payload, the performance during flight, and early results.  

\end{abstract}

\section{Introduction}

The MeV sky is one of the most promising areas of high energy astrophysics that has yet to be adequately explored. Some of the most extreme phenomena in the Universe have their peak emission in the MeV band. However, due to overlapping cross sections between Compton scattering and pair-production for low-MeV gamma-rays in common detector materials\cite{xsec}, building a sensitive MeV instrument is a challenging task. The Imaging Compton Telescope (COMPTEL) \cite{comptel}, de-orbited in 2000, was orders of magnitude less sensitive than instruments observing in neighboring energies, manifesting the so-called ``MeV gap." 

The All-sky Medium Energy Gamma-ray Observatory (AMEGO) \cite{Moiseev:20179z, mcenery2019allsky, Kierans_2020} is a Probe-scale mission concept designed to close the MeV gap in sensitivity. Among AMEGO's science goals are the study of compact objects that also produce neutrinos and gravitational waves, probing element formation in the Universe, and better understanding the physics of astrophysical jets \cite{Moiseev:20179z, mcenery2019allsky}. AMEGO would provide the necessary sensitivity and capability to accomplish these goals. In the mission concept submitted to the Astro2020 Decadal survey, AMEGO consists of a 60 layer Double-sided Silicon Strip Detector (DSSD) tracker, a 3D-position sensitive Cadmium Zinc Telluride (CZT) calorimeter, and a high energy Cesium Iodide (CsI) calorimeter. Surrounding these three subsystems is a plastic scintillator Anti-Coincidence Detector (ACD) for charged particle rejection.    

ComPair is a balloon-borne prototype for AMEGO which tests the viability of the AMEGO design by confirming operation for both Compton and pair-production events, as well as testing the hardware in a space-like environment. ComPair launched from the Columbia Scientific Balloon Facility (CSBF) site in Ft. Sumner, New Mexico, on August 27th, 2023, and completed a 6.5 hour flight reaching an altitude of 133 thousand feet.    

Section 2 of these proceedings describes ComPair's design and instrumentation. Pre-flight testing is covered in Section 3. The balloon flight is detailed in Section 4, and some early results are shown in Section 5. Finally, the next steps for the ComPair/AMEGO project are discussed in Section 6. 

\section{Instrumentation}

\begin{figure}[H]
     \centering
     \begin{subfigure}[t]{0.45\textwidth}
         \centering
         \subcaption{ComPair CAD drawing}
         \includegraphics[height=.7\textwidth]{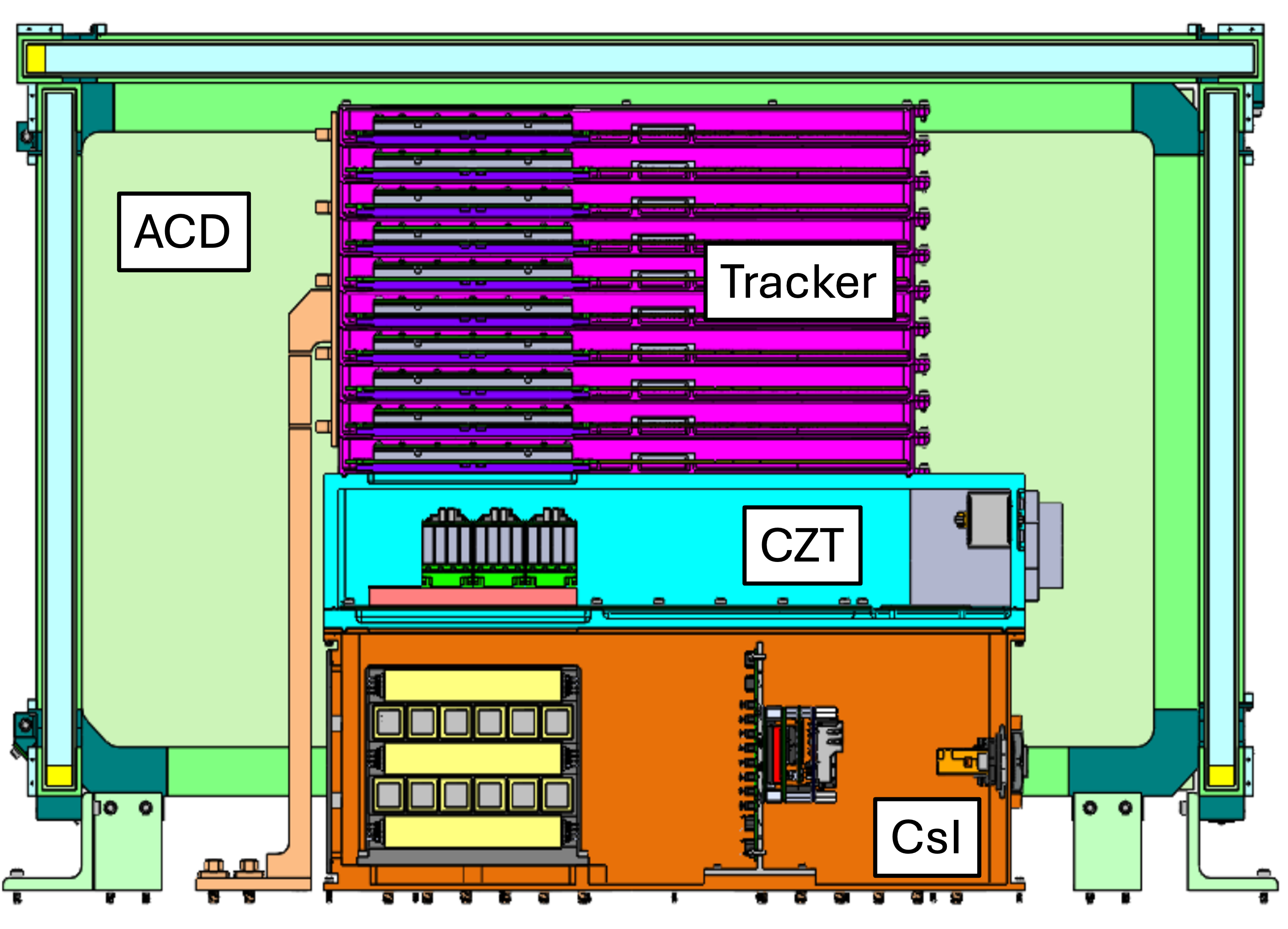}
         \label{fig:cad}
     \end{subfigure}
     \hspace{0mm}
     \begin{subfigure}[t]{0.45\textwidth}
         \centering
         \subcaption{ComPair detector stack with 1 ACD panel removed}
         \includegraphics[height=.7\textwidth]{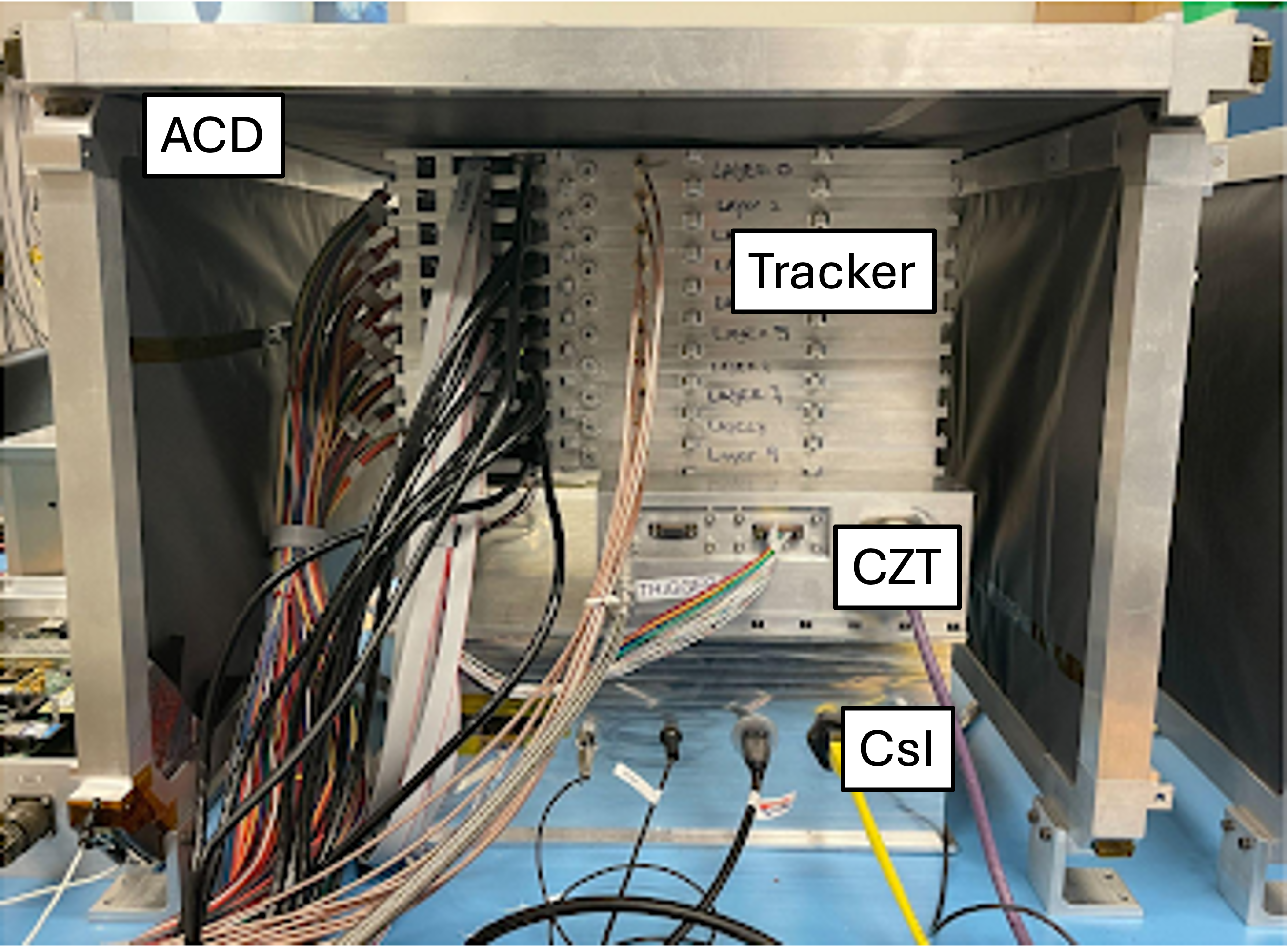}
         \label{fig:subs}
     \end{subfigure}
        \caption{Labeled drawing and photo of the 4 main ComPair subsystems. The detector stack from top to bottom consists of the DSSD Tracker, the CZT calorimeter, and the CsI calorimeter, which are all surrounded by the ACD on 5 sides.}
        \label{fig:instrumentPics}
\end{figure}
 
ComPair consists of 4 main subsystems (Fig. \ref{fig:instrumentPics}): a 10 layer DSSD tracker, a CZT calorimeter, a CsI calorimeter, and a plastic scintillator ACD. The DSSD Tracker is responsible for measuring the positions of charged particles, enabling the tracking of pair-produced positrons and electrons, as well as Compton recoiled electrons. The CZT calorimeter makes  high energy resolution energy measurements of both scattered gamma-rays and pair-produced particles in the low energy range below 10 MeV. The CsI calorimeter covers the high energy response of the instrument. Surrounding the detectors is the plastic scintillator ACD, which vetoes charged particles.

\subsection{Concept of Operation}

\begin{figure}[H]
     \centering
     \begin{subfigure}[t]{0.45\textwidth}
         \centering
         \includegraphics[height=1.0\textwidth]{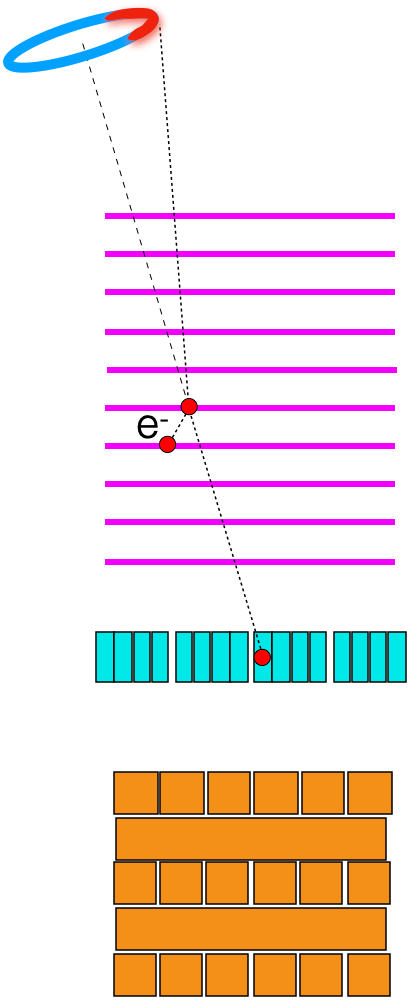}
         \subcaption{Compton scattering: A gamma-ray scatters in the Tracker, and the scattered gamma-ray's energy is measured by a calorimeter. If the the scattered electron is measured, the initial gamma-rays' direction is restricted to an arc (red); if only the scattered gamma-ray is measured, it is restricted to a circle (blue).}
         \label{fig:compton}
     \end{subfigure}
     \hspace{5mm}
     \begin{subfigure}[t]{0.45\textwidth}
         \centering
         \includegraphics[height=1.0\textwidth]{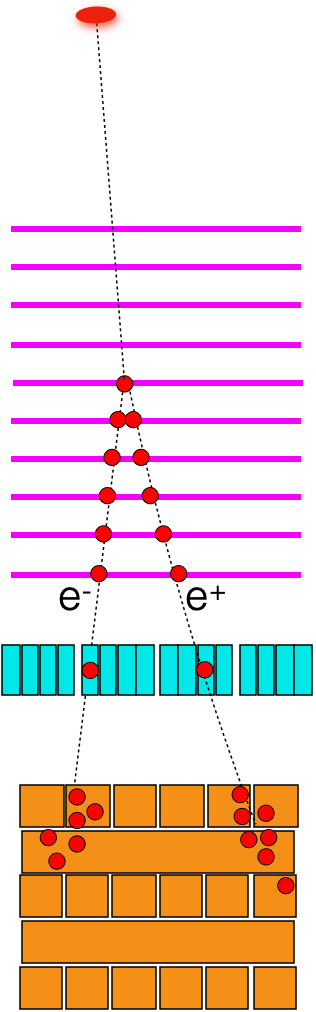}
         \subcaption{Pair-production: A gamma-ray converts into a positron and electron. The pair deposit their energy through each detector, and the direction of the initial gamma-ray can be calculated using conservation of energy and momentum.}
         \label{fig:pair}
     \end{subfigure}
        \caption{Illustrations of Compton scattering and pair-production in the ComPair instrument}
        \label{fig:event_diag}
\end{figure}

ComPair is required to measure and reconstruct both Compton scattering and pair-production events. At lower energies ($<$ $\sim$5 MeV), a gamma-ray will Compton scatter (Fig. \ref{fig:compton}) in the Tracker. The scattered gamma-ray will be absorbed in one of the 2 calorimeters and the path of the recoil electron can be seen in the Tracker. Using the energy of the position of the scatter, and the energy of the scattered gamma ray, the direction of the initial photon can be restricted to an event circle in the sky using the Compton Scattering equation \cite{compton}. Furthermore, if the energy of the recoil electron is measured, the initial direction can be further restricted to an arc on the event circle. For point-like sources, multiple event circles and arcs can be overlapped to localize the source. Above $\sim$5 MeV, pair-production starts to dominate and the majority of gamma-rays will convert into a positron and electron (Fig. \ref{fig:pair}), both of which will then propagate through the tracker and the calorimeters. With the tracks of the positron and electron, and their energy deposited in the calorimeter known, the energy and direction of the gamma-ray can be determined through the kinematics of the pair. 

\subsection{Detectors}

\begin{figure}[t]
\centering
\includegraphics[width=1\linewidth]{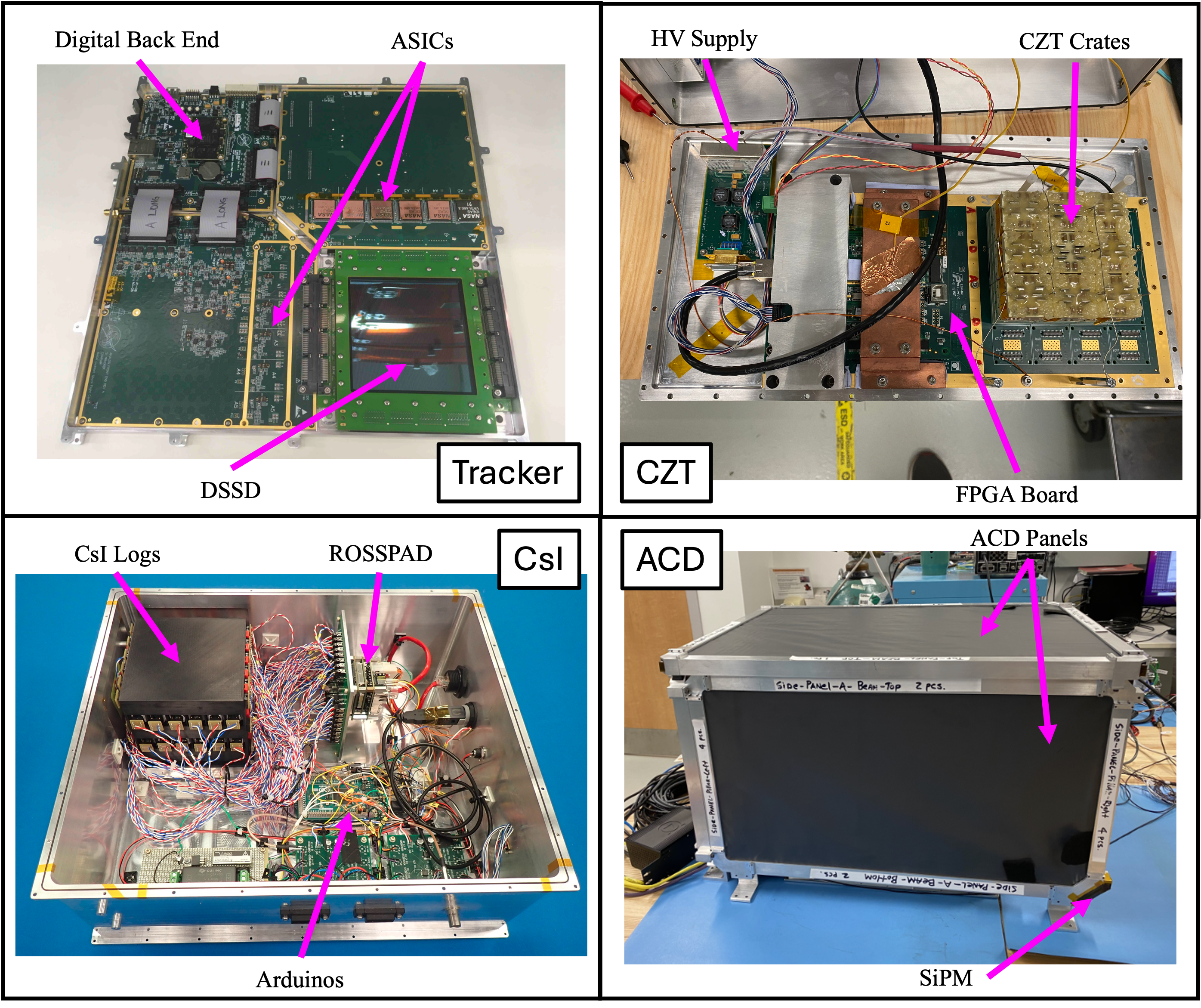}
\caption{\label{fig:internals} Each ComPair subsystem is shown independently. (\textit{Top left}) A single Tracker layer with top cover removed. The DSSD itself is shown in the bottom right, with the ASICs above and to the left, underneath, and the digital back end electronics in the top left. (\textit{Top right}) The CZT calorimeter. Nine crates with 16 CZT bars are shown on the right with 1 ASIC under each crate. The FPGA board is shown in the middle, and the 2500 V bias supply on the left. (\textit{Bottom left}) The CsI calorimeter\cite{Shy_2022}. The 30-log, 5 layer hodoscopic array is shown in the black housing in the top left. The ROSSPAD is shown to the right, and the Arduinos at the bottom. (\textit{Bottom right}) The ACD partially assembled. The plastic scintillator panels are wrapped in 3 layers of black Tedlar, each with a SiPM array at 1 corner.}
\end{figure}

The ComPair Tracker \cite{trackerDevCompair} consists of 10 layers of Double-sided Silicon Strip Detectors (DSSDs) (Fig. \ref{fig:internals}). The detector layers are $10 \times 10 \times 0.05$ cm$^{3}$ with 0.5 mm strip pitch. Each layer has a set of cross-hatched strips to determine the transverse position of interactions. The DSSDs are biased at 60V and readout with IDEAS VATA460.3 ASICs\cite{ideasVata460}. The Tracker has a position resolution 300 $\mu$m (FWHM) and each strip has a dynamic range of 20 - 700 keV, with a measured energy resolution of 4\% ($\sigma$) at 122 keV. 

Below the Tracker is the high resolution Virtual Frisch Grid (VFG) CZT calorimeter\cite{FrischGridCZT} (Fig. \ref{fig:internals}). Consisting of 9 crates of 16 bars each measuring $0.6 \times 0.6 \times 2$ cm$^{3}$, the CZT calorimeter provides high spatial and energy resolution measurements for Compton scattered events. The CZT bars are biased at 2.5 kV and signals are processed and readout using a custom analog ASIC developed by Brookhaven National Laboratory (BNL)\cite{bnlAVG}. The CZT has a dynamic range of 100 keV - 10 MeV, with an energy resolution of 2\% ($\sigma$) at 662 keV, and a 3D position resolution 0.2 cm (FWHM). The CZT bars and front-end electronics are kept in a hermetically sealed pressure vessel to prevent breakout from the high bias voltage in the low pressure balloon environment. 

At the bottom of the instrument stack is the high energy CsI calorimeter, developed by the Naval Research Laboratory (NRL) \cite{Shy_2023} (Fig. \ref{fig:internals}). 30 CsI logs each measuring $1.7 \times 1.7 \times 10$ cm$^{3}$ are arranged hodoscopically in 5 layers. The CsI calorimeter has a dynamic range of 250 keV - 30 MeV per log, with a 2.9\% ($\sigma$) energy resolution at 1.27 MeV, and a position resolution of 1 cm (FWHM). The scintillation light is detected by ONSemi J-Series Silicon photo-multipliers (SiPMs)\cite{onsemiJ} on the end of each bar and read out by an IDEAs 64-channel ROSSPAD\cite{rosspad}. An Arduino Due microcontroller is also used to synchronize CsI events with events in the other subsystems. 

Surrounding the stack is the 5 panel ACD (Fig. \ref{fig:internals}), designed and built by NASA GSFC and NRL. Each panel is 1.5 cm thick EJ 208 plastic scintillator. Wavelength shifting bars along two edges channel the light to ONSemi C-Series SiPMs\cite{onsemiC} on one corner of each panel. A duplicate of the CsI front-end was used for the ACD, which includes the ROSSPAD and Arduino Due. 

\subsection{Balloon Payload Electronics}

\begin{figure}[H]
    \begin{minipage}[t]{.7\textwidth}
        \vspace{0pt}
        \includegraphics[width=1\linewidth]{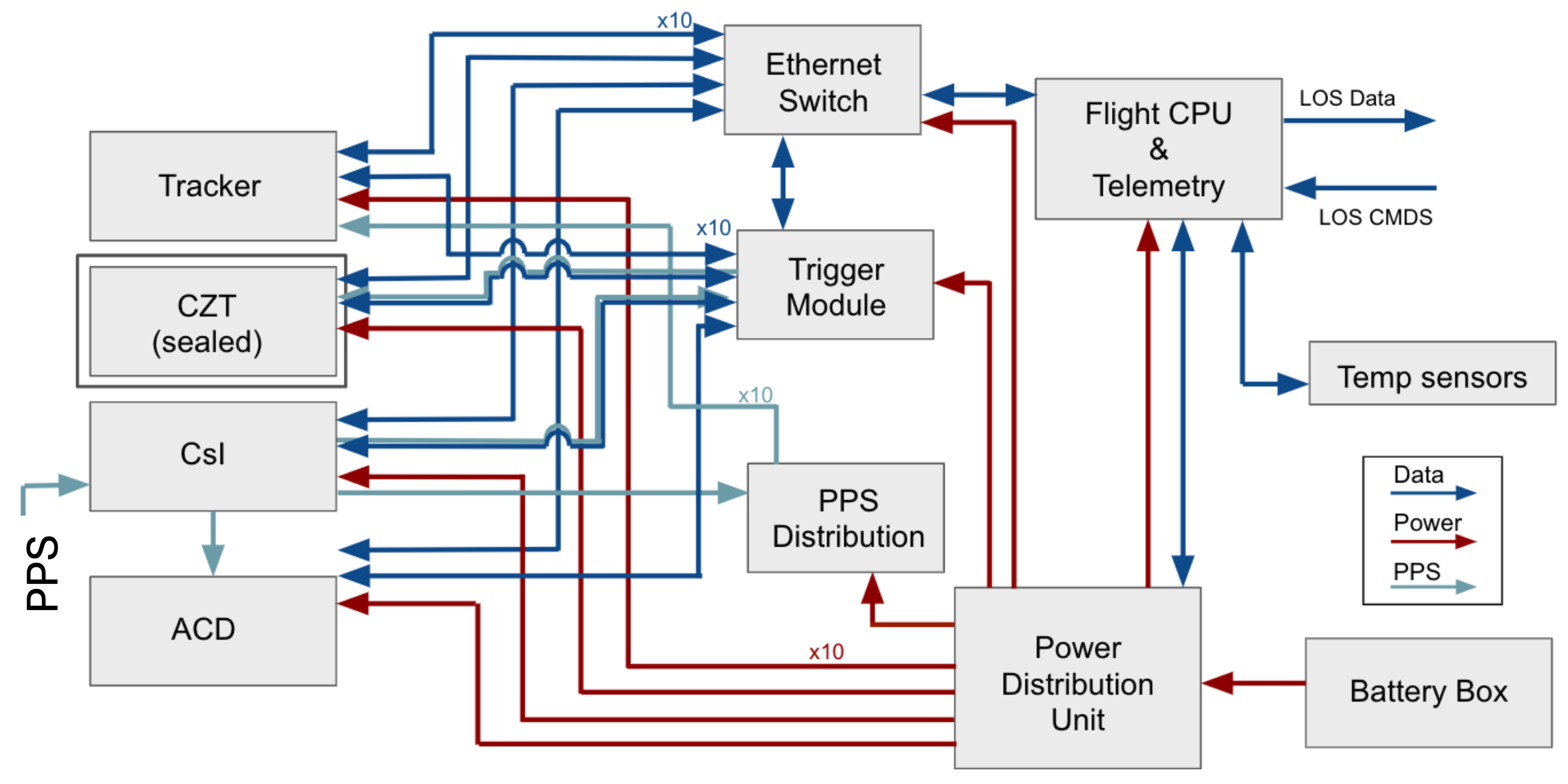}
    \end{minipage}
    \centering
    \hspace{8mm}
    \begin{minipage}[t]{0.2\textwidth}
        \vspace{0pt}
        \caption{\label{fig:block} Block diagram of all ComPair's communication, power, and signal pathways.} 
    \end{minipage}\hfill
\end{figure}

While each subsystem can operate independently through their own DAQ and power systems, they are serviced by several peripheral systems for in-flight operation (Fig. \ref{fig:block}). The Trigger Module (TM) associates hits between each subsystem. A ``hit'' is defined as any interaction in a detector that deposits enough energy to induce a signal greater than the detector-specific threshold. Each subsystem was also capable of recording its own internal triggers in addition to the events acknowledged by the trigger module. The flight computer is responsible for commanding ans data acquisition, and the power distribution unit (PDU) distributes power from the batteries to all other systems. A pulse per second (PPS) signal is read in by the CsI calorimeter from CSBF hardware. It is then sent to the PPS distribution module, which routes it to the other detectors for internal clock correction.

\subsubsection{Trigger Module}

The ComPair Trigger Module (Fig \ref{fig:tm}) \cite{MakotoTM} checks for coincidences in hits seen in the subsystems. The TM is able to check for multiple, preset coincidence modes, where the data for certain combinations of hits in each detector is read out. The TM receives the internal hit signals from each detector and sends a trigger acknowledge signal back if one of the coincidence modes is satisfied. If so, each subsystem records the data for that event. An event identification number (EventID) is also generated for each acknowledged event to associate events between each subsystem. The four chosen hit conditions for the balloon flight required a hit in: (1) both sides of a single layer, (2) at least one tracker layer and the CZT calorimeter, (3) at least one tracker layer and the CsI calorimeter, and (4) the CZT calorimeter and CsI calorimeter. 

\subsubsection{Data Acquisition \& Flight computer}

\begin{figure}[H]
     \centering
     \begin{subfigure}[t]{0.45\textwidth}
         \centering
         \includegraphics[height=.7\textwidth]{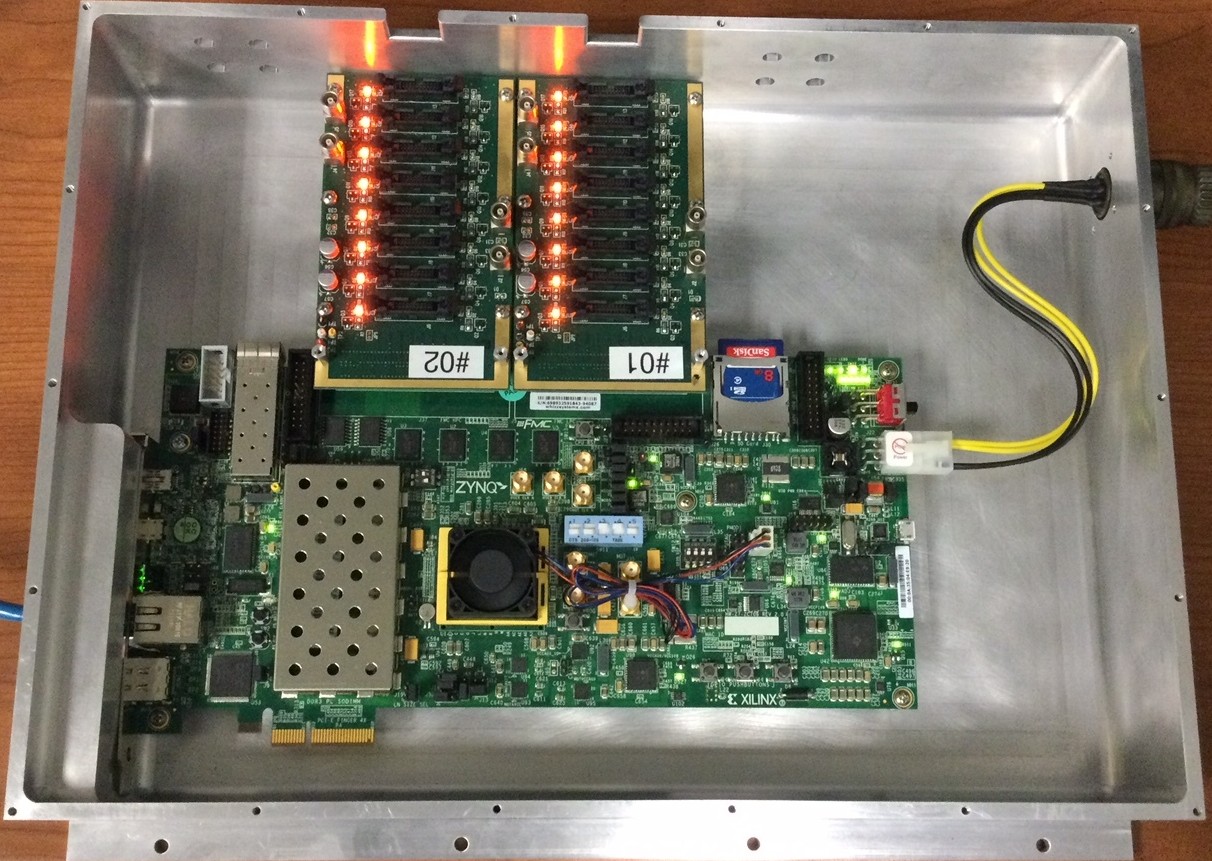}
         \subcaption{The ComPair TM, consisting of a  Xilinx ZC706 FPGA evaluation board (bottom) and 2 connection boards (top). Each Tracker layer, the CZT, the CsI, and ACD, have a cable which connects to the TM through a port on one of the two connection boards.}
         \label{fig:tm}
     \end{subfigure}
     \hspace{5mm}
     \begin{subfigure}[t]{0.45\textwidth}
         \centering
         \includegraphics[height=.7\textwidth]{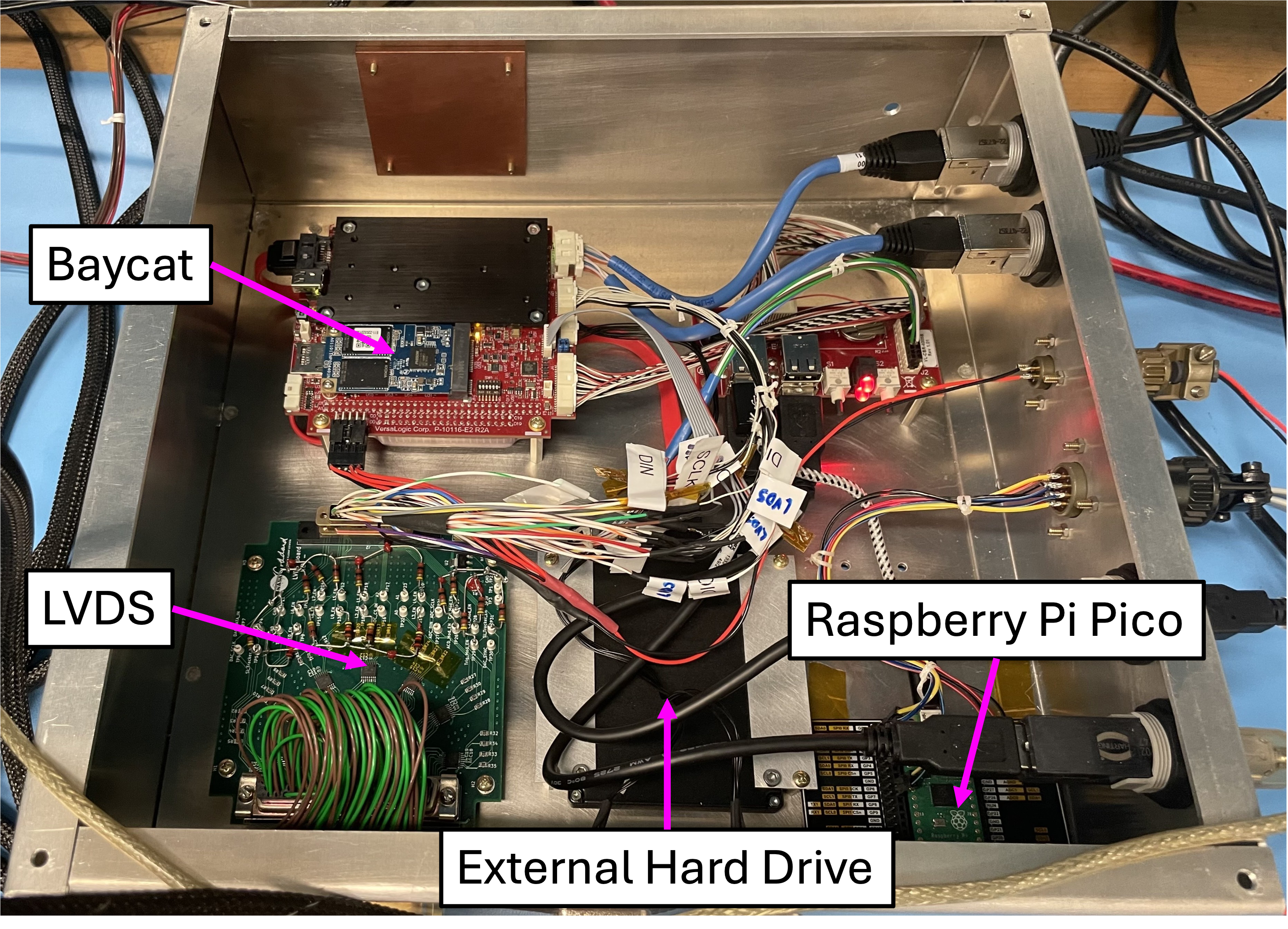}
         \subcaption{The ComPair CPU enclosure: the Baycat SBC sits at the top right, the low voltage differential signalling (LVDS) board, which is responsible for enabling/disabling lines in the PDU, sits bottom left. The external hard drive sits on the bottom center, and a Raspberry Pi Pico microcontroller, used for temperature monitoring, sits bottom right.}
         \label{fig:cpu}
     \end{subfigure}
        \caption{Support electronics for the ComPair balloon flight: the Trigger Module (TM) and flight }
        \label{fig:periph}
\end{figure}

ComPair uses a Versalogic BayCat single board computer (SBC) as the flight computer, or Central Processing Unit (CPU, Fig. \ref{fig:cpu}), which is responsible for the data acquisition and in-flight commanding for all other systems. The CPU software was developed by Los Alamos National Laboratory (LANL), which uses an Elixir Open Telecom Platform application to perform data collection, control power distribution, and readout thermometry. The CPU connects to the CSBF miniature Science Instrument Package (SIP) to downlink housekeeping and data during the flight, as well as two external hard drives for redundant data storage. The CPU communicates with the detectors and the TM through an Antaira LMX-1802G-SFP Ethernet switch\cite{switch}. 

\subsubsection{Power Distribution \& Thermometry}

An array of eight 28 V lead-acid batteries in parallel powered all systems during the balloon through the power distribution unit (PDU). As a whole ComPair uses 290 W of power, while the Tracker, CZT, CsI, and ACD uses 95, 13, 6, and 6 W, respectively. Due to the considerable heat dissipation from the power consumed by the Tracker, copper straps were machined to conduct heat to the baseplate to radiate away as the instrument is enclosed by the ACD. ComPair used a total of 18 MCP9808\cite{mcp} digital temperature sensors placed throughout the instrument which monitored heat dissipating during thermal vacuum tests and the balloon flight.

\subsection{Data Pipeline \& Analysis Methods}

Each subsystem has its own data acquisition system that runs independently generates binary data files, which are then converted to Hierarchical Data Format (HDF) files. A series of subsystem-specific calibrations are applied and the separate detector data are combined into a single data format, the ComPair Unified Data (CUD) format, which also uses HDF structure. 

ComPair makes extensive use of the Medium-Energy Gamma-ray Astronomy Library (MEGAlib)\cite{megalib} for data analysis and simulation. Revan (real event analyzer) and Mimrec (Megalib Image REConstruction) are used for event reconstruction and imaging. CUD level data files are converted to Revan-readable files and then high-level analysis can be done with Revan and Mimrec, and Monte Carlo simulations are performed with the Geant4-based Cosima (Cosmic simulator) tool.  

The detector effects engine (DEE) is a custom built python based tool which takes in simulated data from Cosima and applies an inverse calibration to raw binary data. The simulated detector level data is then run through the standard ComPair calibration pipeline, which allows us to investigate possible instrumental effects in real data.

\section{Pre-flight Performance Validation}

ComPair underwent environmental testing to ensure detector performance and thermal viability in the months leading up to the launch, both in the lab at NASA GSFC and on site in Fort Sumner. Calibrations with radioactive sources verified the energy and spatial response of the detectors towards the low energy side of ComPair's dynamic range, while a beam test covered the high energy regime. Thermal-vacuum (TVAC) tests ensured operation and survivability of the detectors and all other systems in balloon altitude pressure and extreme temperature environments. 

\subsection{Calibrations}

\begin{table}[h]
  \begin{minipage}[t]{0.67\textwidth}
    \vspace{0pt}
    \begin{tabular}{|c|c|c|c|}
        \hline
        Source & Activity (Ci) & Energy (MeV) & Branching Ratios (\%)\\
        \hline
        $\prescript{133}{}{\text{Ba}}$ & 7.93e-06 & 0.303, 0.356 & 18, 62  \\
        $\prescript{22}{}{\text{Na}}$ & 4.64e-04 & 0.511, 1.27 & 180, 100  \\
        $\prescript{137}{}{\text{Cs}}$ & 6.39e-04 & 0.662 & 85  \\
        $\prescript{60}{}{\text{Co}}$ & 5.54e-04 & 1.17, 1.33 & 99, 100 \\
        \hline
    \end{tabular}
  \end{minipage}
  \centering
  \begin{minipage}[t]{0.3\textwidth}
    \vspace{0pt}
    \caption{A list of the radioactive sources ComPair used for calibrations, the activity of the source, and the energy and branching ratios of their decay products.}
    \label{tab:sources}
  \end{minipage}\hfill
\end{table}

\begin{figure}[H]
     \centering
     \begin{subfigure}[t]{0.45\textwidth}
         \centering
         \includegraphics[height=.8\textwidth]{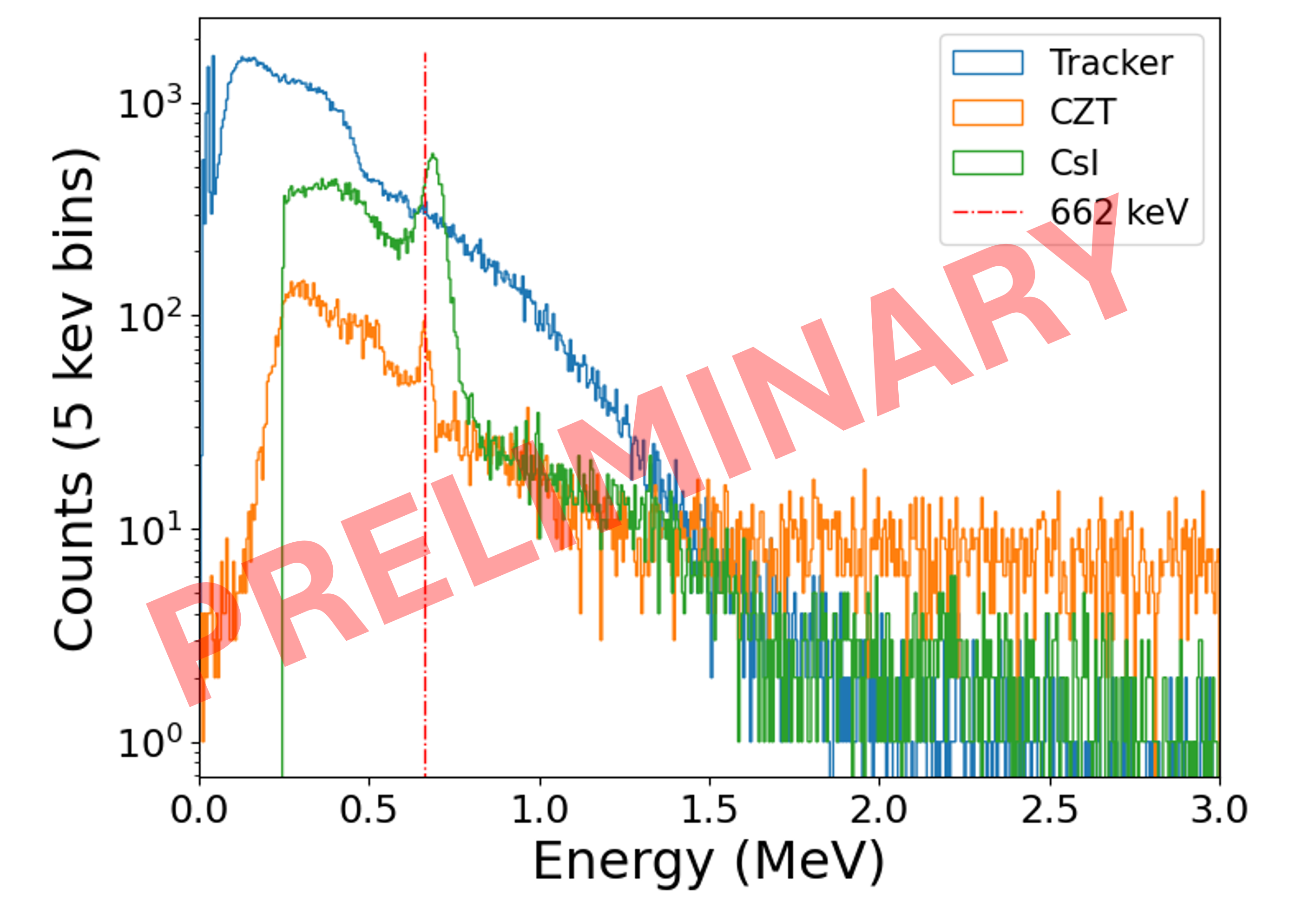}
         \subcaption{Individual subsystem energy spectra for $\prescript{137}{}{\text{Cs}}$. The 662 keV photopeak is visible in both the CZT and CsI spectra.}
         \label{fig:cs137_ind}
     \end{subfigure}
     \hspace{10mm}
     \begin{subfigure}[t]{0.45\textwidth}
         \centering
         \includegraphics[height=.8\textwidth]{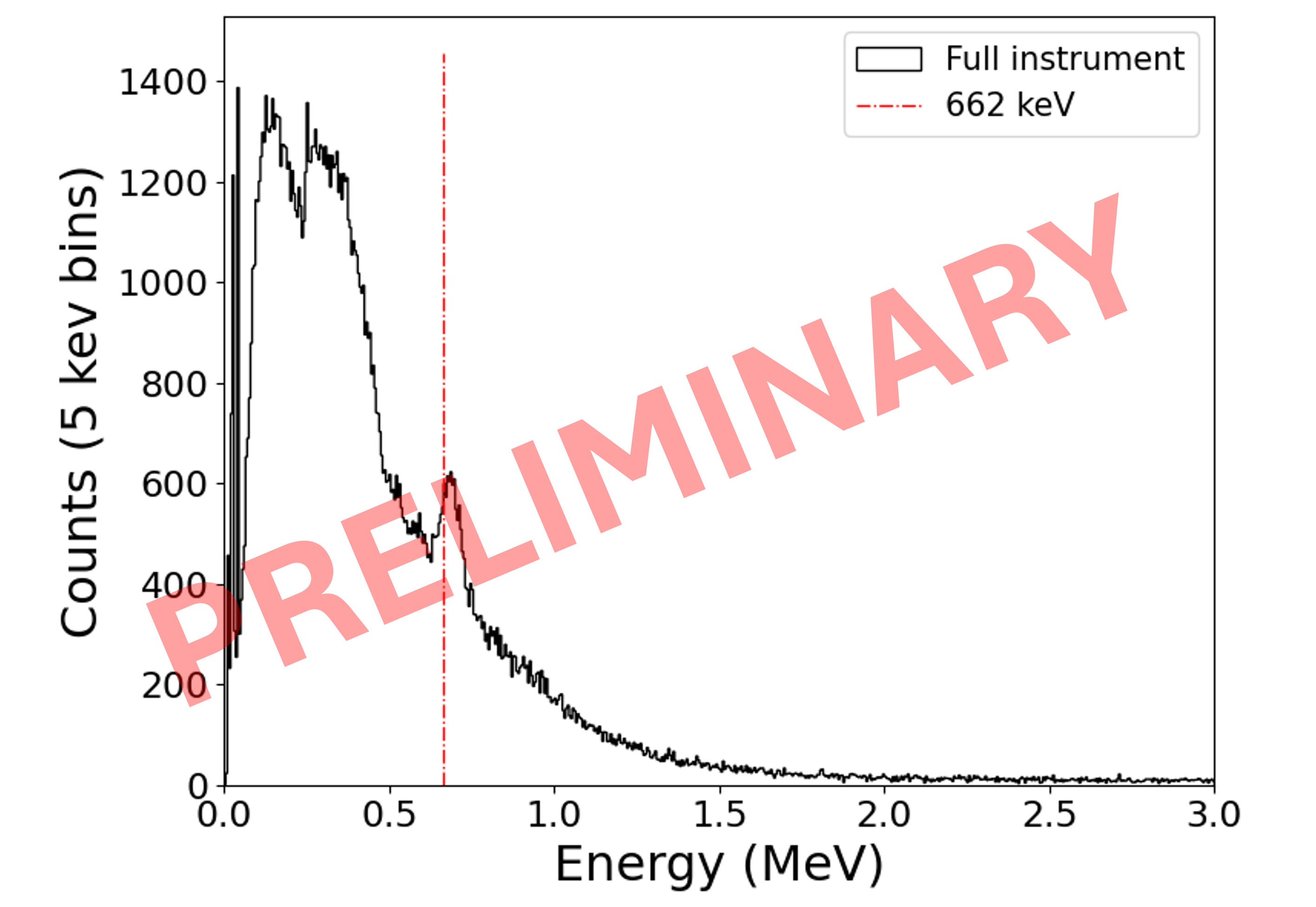}
         \subcaption{Full instrument $\prescript{137}{}{\text{Cs}}$, where energies from the 3 subsystems are summed event by event.}
         \label{fig:cs137_com}
     \end{subfigure}
        \caption{Preliminary $\prescript{137}{}{\text{Cs}}$ spectra.}
        \label{fig:cs137}
\end{figure}

Four radioactive sources (see Table \ref{tab:sources}) were used in calibrating the instrument, with energies ranging from 0.303 MeV to 1.33 MeV. The sources were placed several meters from the instrument depending on their specific activity, and at polar angles ranging from 0$^\circ$ to 50$^\circ$ to understand the instrument response from different source positions. Fig. \ref{fig:cs137} shows an example energy spectra from a $\prescript{137}{}{\text{Cs}}$ source for all 3 detectors, as well combined on an event by event basis for the whole instrument. The goal of these tests is to measure ComPair's full system energy resolution, angular resolution, and effective area.

\begin{figure}[H]
\centering
\includegraphics[width=0.5\linewidth]{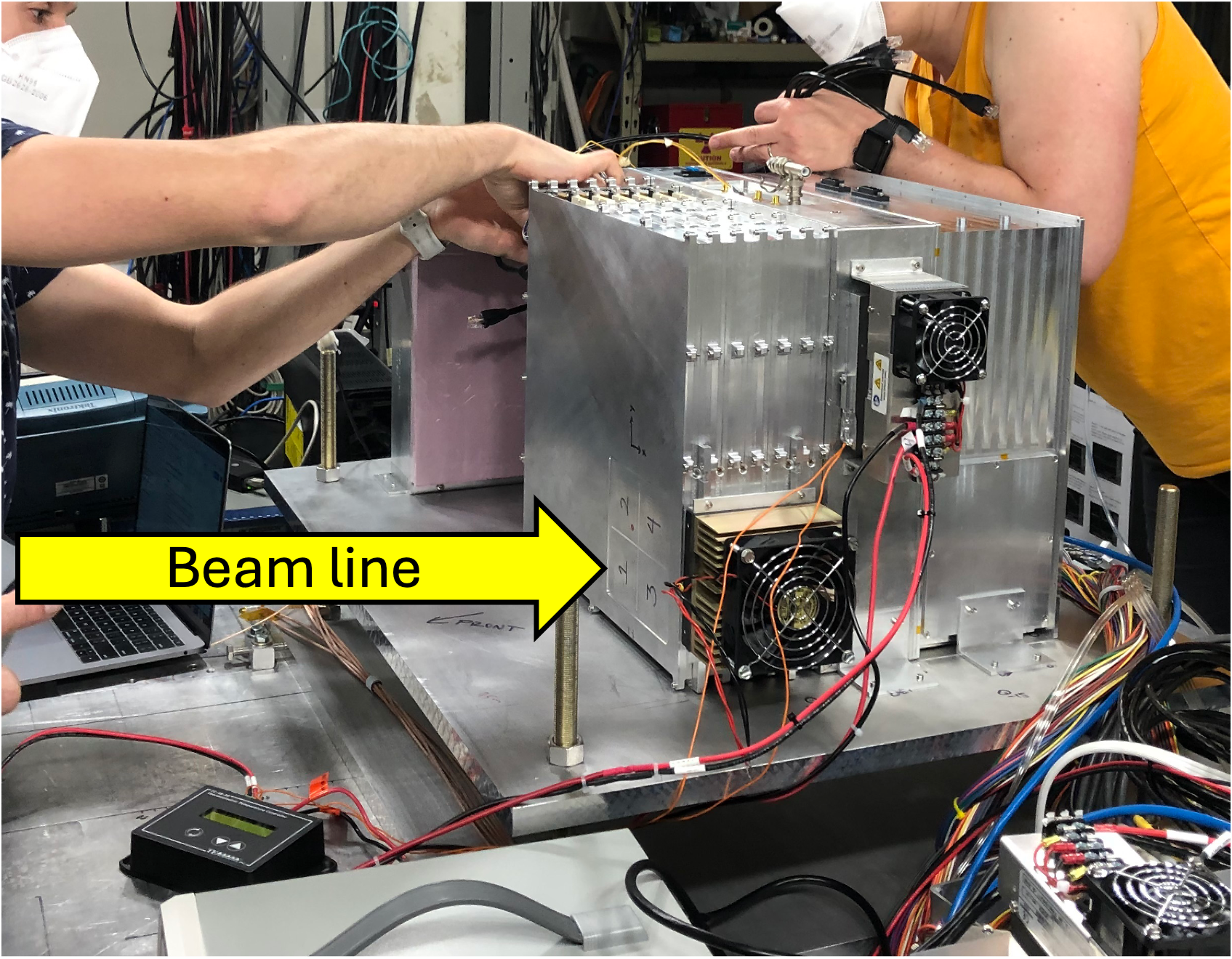}
\caption{\label{fig:higs} ComPair being assembled at the HIGS beam test. Five tracker layers were available at the time of the test. The gamma-ray beam enters from the left, so the Tracker, CZT, and CsI are placed on their side.}
\end{figure}

ComPair was also tested at the Triangle Universities Nuclear Lab's High Intensity Gamma-ray Source (HIGS)\cite{higs} in April 2022 (Fig. \ref{fig:higs}). HIGS provided a collimated gamma-ray beam at from energies 2 MeV up to 25 MeV, which is mostly unachievable with common radioactive sources. This energy range allows ComPair to probe both the Compton scattering and pair-production regimes, as well as their cross-over region ($\sim$5 MeV). Five Tracker layers, the CZT, and the CsI, and the Trigger module successfully operated during the test. The calibration analysis is currently in progress, but preliminary results from the HIGS beam test can be found in \citen{Shy_2022}. 

\subsection{TVAC}

Each subsystem was individually tested in a TVAC chamber to validate their performance in a balloon-like environment. The integrated instrument with the flight periphery was tested at balloon pressures and temperatures in a large chamber at GSFC to measure the instrument response in these conditions. These results were compared with a thermal model of the instrument to see if the predictions matched the data during flight. With the detectors and electronics powered off, the chamber was pumped down to 6 Torr and the chamber shroud was set to -20$^\circ$C. 

\begin{figure}[H]
\centering
\includegraphics[width=1\linewidth]{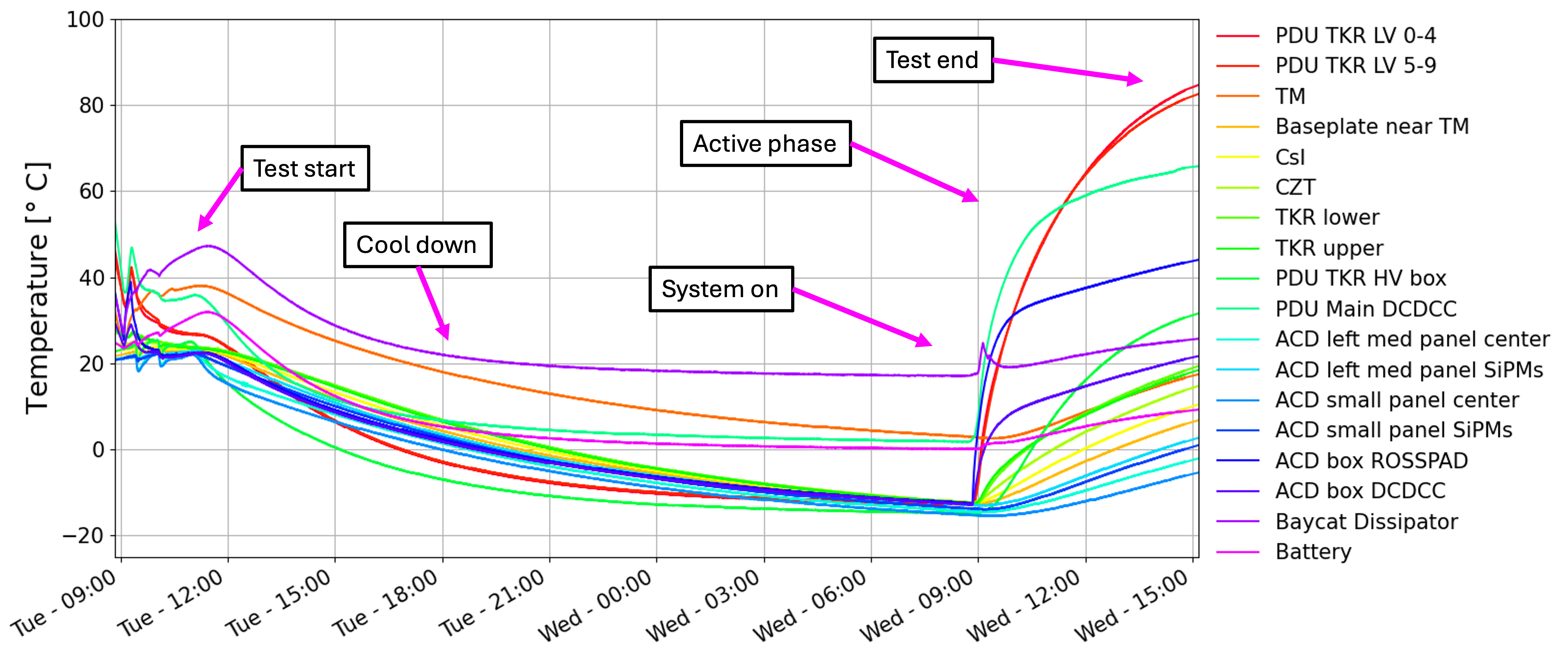}
\caption{\label{fig:tvac} TVAC test temperature profiles from all thermometers. All systems were powered off at the start of the test, and allowed to cool down for 24 hours to equilibrate with the chamber, set at -20$^\circ$C. Most systems reached around -15$^\circ$C, while the TM, PDU DCDC converter, and Battery reached about 0$^\circ$C, and the CPU dissipator at 17$^\circ$C. The active phase of the test started once the system was powered on, and lasted for 6 hours.}
\end{figure}

After allowing 24 hours to cool down, most subsystems equilibrated around -16$^\circ$C. The system was powered on and began self-heating (Fig. \ref{fig:tvac}). The temperature increases were used to inform thermal models for performance during flight. Due to high temperatures reached in the PDU, greater than 80$^\circ$C, a copper heat sink was machined to mitigate over-heating during flight.

\section{Balloon Flight \& Preliminary Results}

\subsection{Balloon Gondola}

\begin{figure}
\centering
\includegraphics[width=0.5\linewidth]{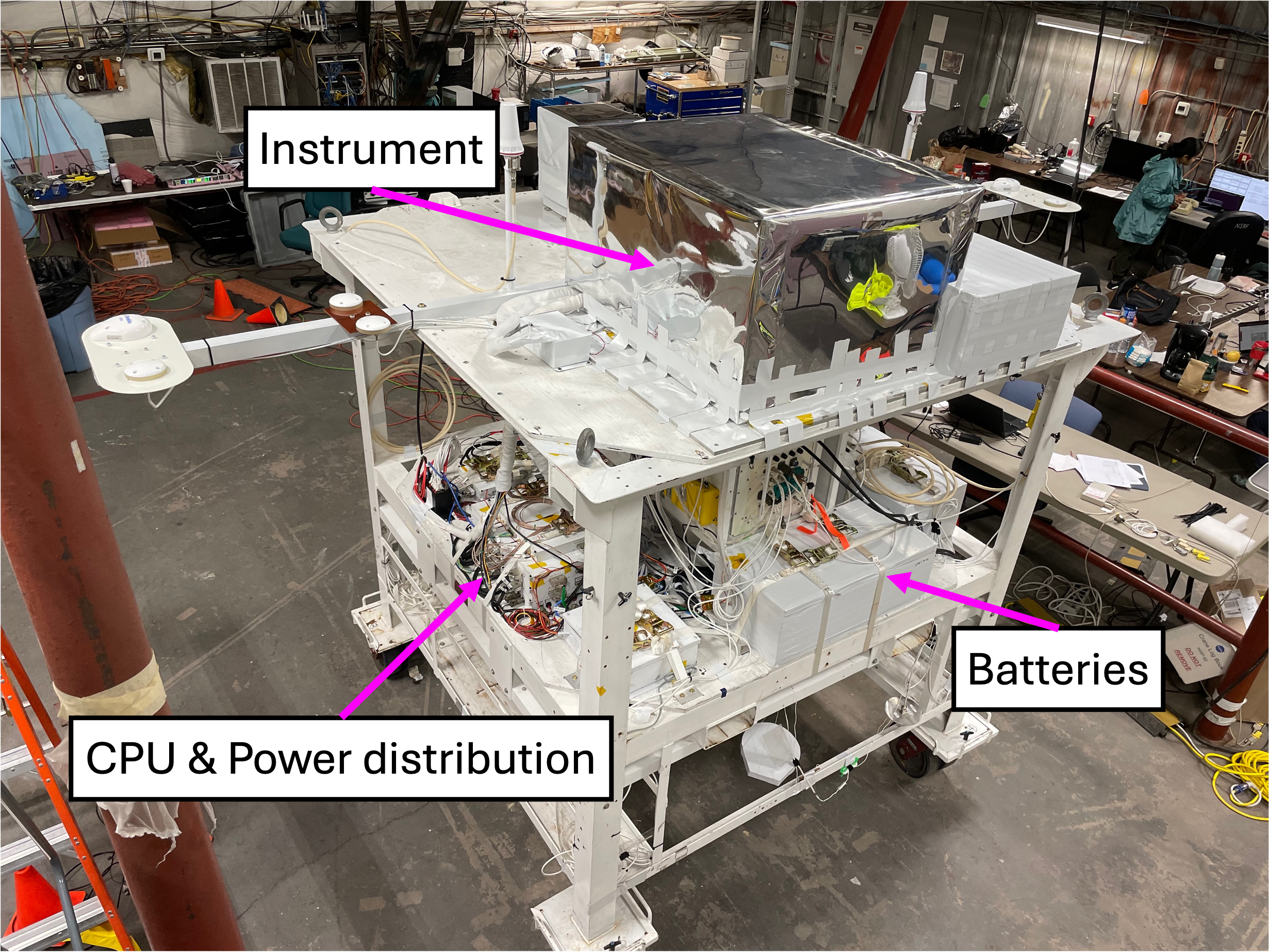}
\caption{\label{fig:gond} ComPair fully integrated on the Iron Maiden gondola. The instrument, TM, PPS distribution, and ACD front-end sit on the top level, and the CPU, PDU, Ethernet switch, and batteries on the middle level. }
\end{figure}

ComPair was a piggy back instrument to the Gamma-RAy Polarimeter Experiment (GRAPE)\cite{grape}, both of which were mounted on the CSBF Iron Maiden gondola (Fig. \ref{fig:gond}). The gondola did not require any poitning capability, thus was able to freely rotate during the flight. The detector stack was placed on the top level, allowing an unobstructed view of the sky, along with the Trigger Module, the ACD front-end electronics, and the PPS distribution module.  The CPU, Power Distribution, Ethernet switch, and battery arrays were all placed on the middle level. All of the boxes on the top level were insulated with a foam box, which was then covered in either a layer of aluminized mylar (the detectors) or white duct tape (TM, ACD electronics). The balloon was a zero-pressure 39 million ft$^{3}$, 0.8 mil thick balloon. 

\subsection{Flight Details}

\begin{figure}[H]
     \centering
     \begin{subfigure}[t]{0.45\textwidth}
         \centering
         \includegraphics[height=.7\textwidth]{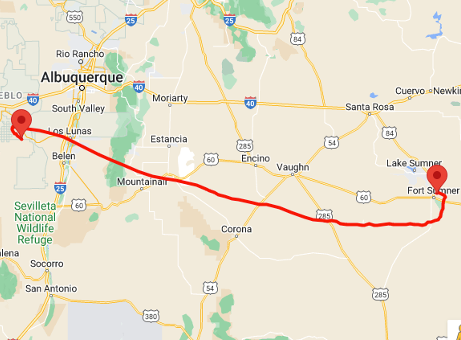}
         \subcaption{ComPair's flight trajectory, launching from Fort Sumner, NM and landing southwest of Albuquerque, NM.}
         \label{fig:traj}
     \end{subfigure}
     \hspace{5mm}
     \begin{subfigure}[t]{0.45\textwidth}
         \centering
         \includegraphics[height=.7\textwidth]{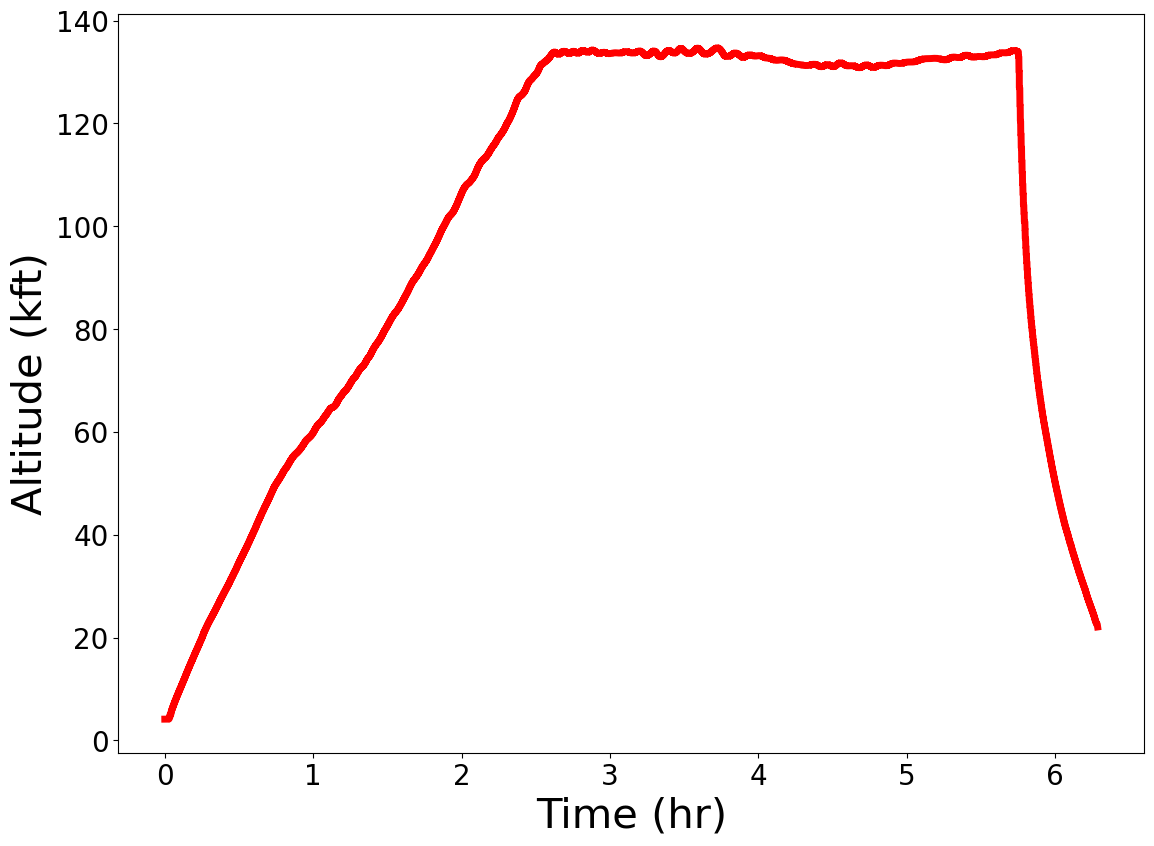}
         \subcaption{Altitude vs. time profile for the balloon flight.}
         \label{fig:alt}
     \end{subfigure}
        \caption{}
        \label{fig:Trajectory & altitude profile}
\end{figure}

ComPair launched around 9am MDT, from Fort Sumner, NM (\ref{fig:traj}). A floating altitude of 130,000 feet was reached in approximately 3 hours, and remained at float for 3 hours (\ref{fig:alt}). The payload descended for about 30 minutes and landed southwest of Albuquerque, NM, for a roughly 6.5 hour total flight time. The detectors were powered on about 10 minutes after launch to measure the radiation environment during the ascent. At both 4 and 5 hours after launch, the detectors were powered down for roughly 15 minutes, due to high temperatures reached in some of the subsystems, then restarted with no issue. After the flight it was discovered that both of the Arduinos in the CsI and ACD failed during the first few minutes of the flight. The consequence of this being that events in both systems were not automatically synchronized with the other subsystems. This problem was solved by associating the PPS signal data between the CsI and the TM, and matching the EventIDs from the TM data to hits in the CsI and ACD. 

For most of the flight, ComPair used a soft ACD veto, meaning that all events satisfying triggering conditions were recorded, regardless of a detection in the ACD. A hard veto, where events with coincidence triggers in the ACD were rejected, was enabled during the 12 minutes of the flight, which is reflected in the drop in event rate in Fig . Data acquisition periods were set to 30 minutes, resulting in periodic event losses as the CPU switches to writing new data files. 

\begin{figure}[H]
\centering
\includegraphics[width=1.0\linewidth]{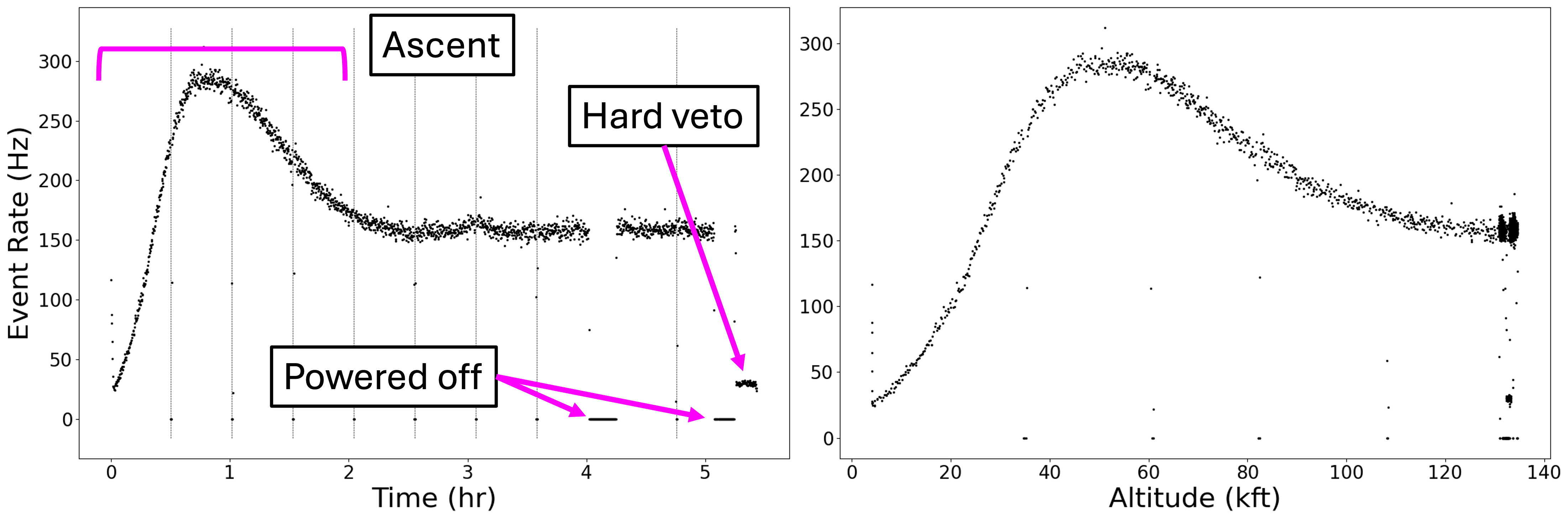}
\caption{ Event rate as a function of time (\textit{left}) and altitude (\textit{right}). The vertical lines on the left plot denote the 30 minute data cycle. The Regener-Pfotzer Maximum is visible in each plot during the ascent, at 45 minutes and 55 kft, respectively. The instruments were powered off at 4 and 5 hours, for about 15 minutes to allow for cooling. The hard ACD veto used near the end of the flight is reflected in the decrease of the event rate in the left plot.}
\label{fig:evt_rt}
\end{figure}

\subsection{Preliminary Flight Results}

The total ComPair event rate is plotted in Fig. \ref{fig:evt_rt}. These events are dominated by charged particles produced in cosmic ray air shower interactions in the atmosphere. The event rate increases from the ground level background up to a maximum near 300 Hz, as the payload moves through the region of maximal radiation from cosmic ray interactions, the Regener-Pfotzer Maximum\cite{REGENER1935}, near 55 kft, and then decreases to around 160 Hz when floating altitude is reached.

\begin{figure}[H]
     \centering
     \begin{subfigure}[t]{0.49\textwidth}
         \centering
         \includegraphics[height=.7\textwidth]{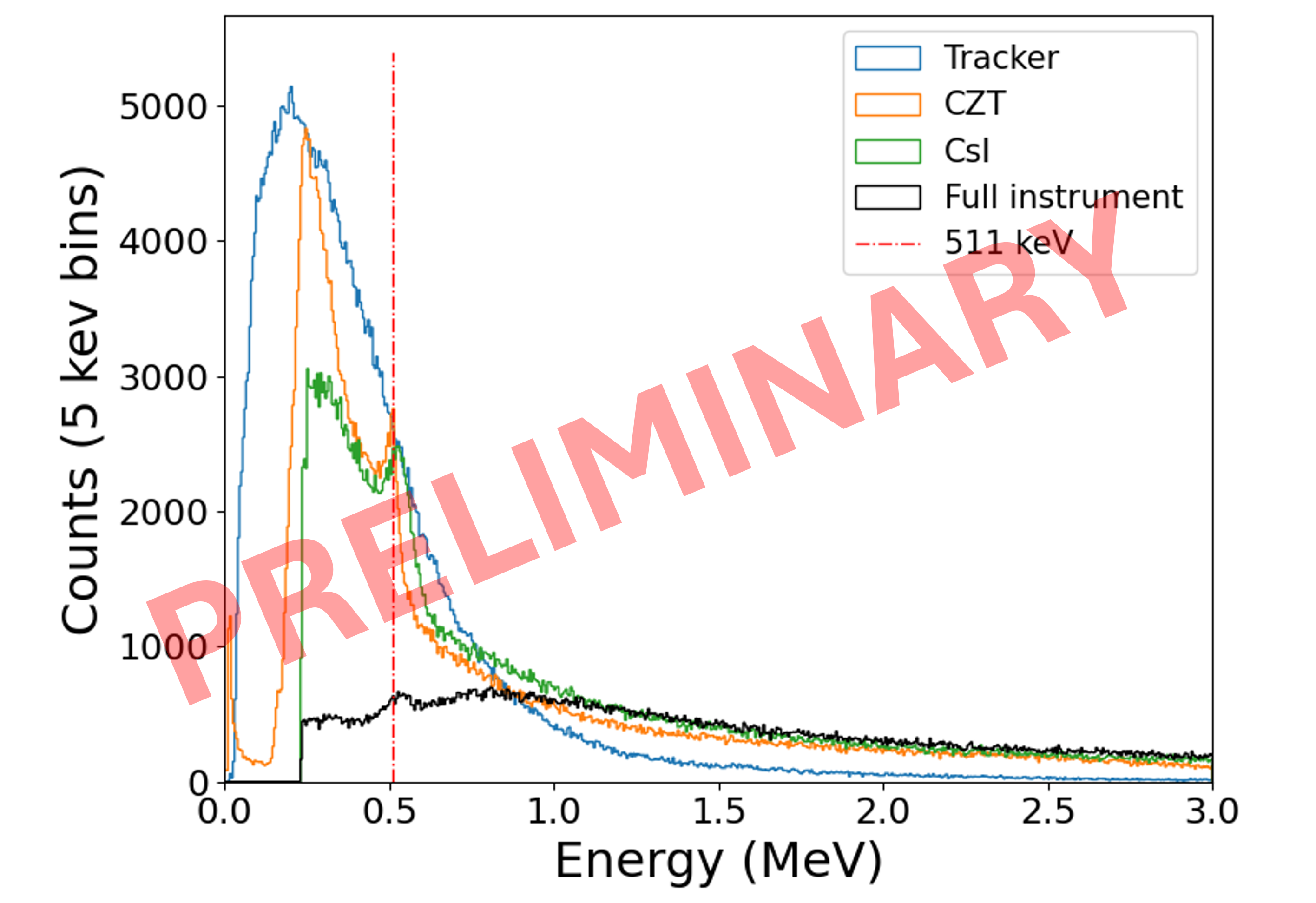}
         \subcaption{Energy spectra from the balloon plotted plotted from 0 to 3 MeV. The 511 keV postritron annihilation line in visible in the CZT, CsI, and full instrument spectra.}
         \label{fig:flightspec1}
     \end{subfigure}
     \hspace{0mm}
     \begin{subfigure}[t]{0.49\textwidth}
         \centering
         \includegraphics[height=.7\textwidth]{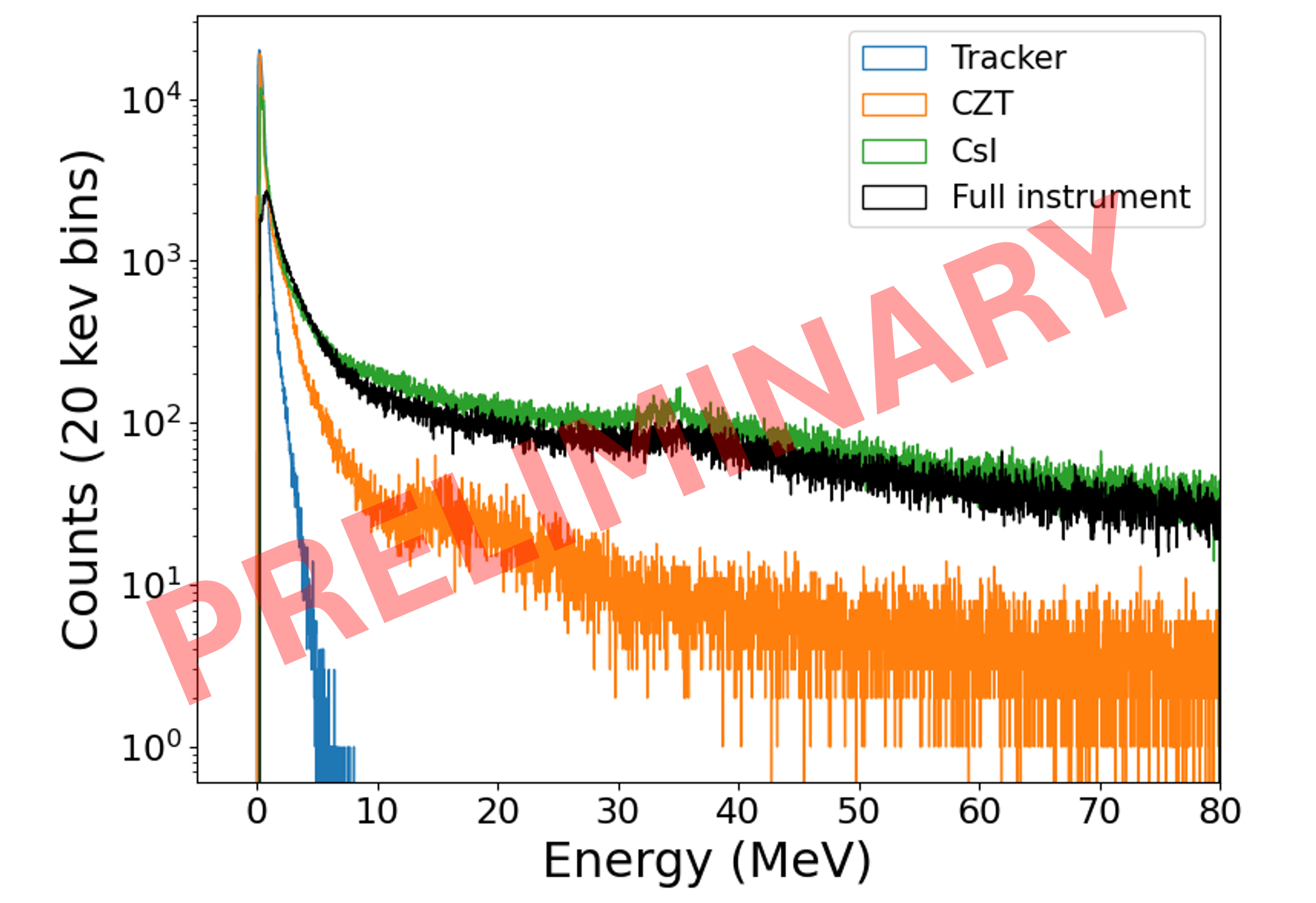}
         \subcaption{Energy spectra from the balloon flight plotted from 0 to 80 MeV. The possible feature near 35 MeV in the CsI and full instrument spectra is currently being investigated.}
         \label{fig:flightspec2}
     \end{subfigure}
        \caption{Preliminary spectra from the balloon flight, for individual subsystems and combined as a whole instrument.}
        \label{fig:flightspec}
\end{figure}

The energy spectra from the balloon flight for each detector, and combined as a whole instrument, summed for each event is shown in Fig \ref{fig:flightspec}. The atmospheric positron annihilation line at 511 keV is observed in the CZT, CsI, and combined spectra. The full spectrum peaks around 1 MeV, which is believed to be where ComPair's effective area is the greatest. Further analysis is in progress, including matching the measured background spectra to background simulations, and to better understand the features at $\sim$17 MeV in the CZT and $\sim$35 MeV in the CsI. One possible explanation is the peak energy deposition from Minimally Ionizing Particles (MIPs) in each detector. More detailed results from the Tracker, CsI, and ACD can be found in \citen{kirschy}, \citen{metzler}, \citen{shy24}. 

\section{Conclusions and Future work}

ComPair achieved its goal of all systems remaining operational during the flight, and measured gamma-ray backgrounds at balloon-flight altitude. The Regener-Pfotzer maximum and the 511 keV positron annihilation line were both observed during the flight. Preliminary results from the balloon flight have been presented, but the full-instrument calibrations, and in turn energy resolution, effective area, and angular resolution measurements are ongoing. The next iteration of ComPair, ComPair 2\cite{compair2} will serve as a technology demonstration balloon flight for the All-sky Medium Energy Gamma-ray Observatory eXplorer, AMEGO-x\cite{AMEGOX}, a medium class explorer mission.

\section{Acknowledgements}

The material is based upon work supported by NASA under award number 80GSFC21M0002. This work was sponsored by NASA Astrophysics and Research Analysis (APRA) grants NNH21ZDA001N-APRA, NNH18ZDA001N-APRA, NNH14ZDA001N-APRA,
and NNH15ZDA001N-APRA. Daniel Shy is supported
by the U.S. Naval Research Laboratory’s Jerome and Isabella Karle Distinguished Scholar Fellowship Program. The authors are thankful to NASA CSBF and to the GRAPE team for accommodating the ComPair instrument.

\bibliography{main} 

\begin{thebibliography}{10}

\bibitem{xsec}
``{NIST XCOM}.'' https://physics.nist.gov/PhysRefData/Xcom/html/xcom1.html.

\bibitem{comptel}
Schoenfelder, V., ``Instrument description and performance of the imaging gamma-ray telescope comptel aboard the compton gamma-ray observatory,'' {\em ApJS}  (1993).

\bibitem{Moiseev:20179z}
Moiseev, A., ``{All-sky Medium Energy Gamma-ray Observatory (AMEGO)},'' {\em PoS}~{\bf ICRC2017},  798 (2017).

\bibitem{mcenery2019allsky}
McEnery, J., ``{All-sky Medium Energy Gamma-ray Observatory: Exploring the Extreme Multimessenger Universe},'' (2019).

\bibitem{Kierans_2020}
Kierans, C.~A., ``{AMEGO: exploring the extreme multi-messenger universe},'' in [{\em Space Telescopes and Instrumentation 2020: Ultraviolet to Gamma Ray}{\nolinebreak\hspace{0.1em}]},  den Herder, J.-W.~A., Nakazawa, K., and Nikzad, S., eds., SPIE (Dec. 2020).

\bibitem{compton}
Compton, A.~H., ``A quantum theory of the scattering of x-rays by light elements,'' {\em Phys. Rev.}~{\bf 21},  483--502 (May 1923).

\bibitem{Shy_2022}
Shy, D., Kierans, C.~A., Cannady, N., Caputo, R., Griffin, S., Grove, E., Hays, E., Kong, E., Kirschner, N., Liceaga-Indart, I., McEnery, J.~E., Mitchell, J., Moiseev, A., Parker, L., Perkins, J., Phlips, B., Sasaki, M., Schoenwald, A.~J., Sleator, C., Smith, J.~R., Smith, L., Wasti, S., Woolf, R.~S., Wulf, E., and Zajczyk, A., ``Development of the compair gamma-ray telescope prototype,'' in [{\em Space Telescopes and Instrumentation 2022: Ultraviolet to Gamma Ray}{\nolinebreak\hspace{0.1em}]},  den Herder, J.-W.~A., Nakazawa, K., and Nikzad, S., eds., SPIE (Aug. 2022).

\bibitem{trackerDevCompair}
Griffin, S., Kierans, C., Parker, L., Schoenwald, A., Shawhan, P., Caputo, R., McEnery, J., and Perkins, J., ``{Current status of the ComPair silicon tracker},'' in [{\em Space Telescopes and Instrumentation 2020: Ultraviolet to Gamma Ray}{\nolinebreak\hspace{0.1em}]},  den Herder, J.-W.~A., Nikzad, S., and Nakazawa, K., eds.,  {\bf 11444},  547 -- 555, International Society for Optics and Photonics, SPIE (2020).

\bibitem{ideasVata460}
``{Vata460.3 ideas}.'' https://ideas.no/products/vata460-3/ (2022).

\bibitem{FrischGridCZT}
Hays, E.~A., Bolotnikov, A., Kierans, C., Moiseev, A., and Thompson, D., ``{{Modular Position-sensitive High-resolution Calorimeter for Use in Space Gamma-ray Instruments Based on Virtual Frisch-grid CdZnTe Detectors}},'' {\em PoS}~{\bf ICRC2019},  584 (2020).

\bibitem{bnlAVG}
Vernon, E., {De Geronimo}, G., Bolotnikov, A., Stanacevic, M., Fried, J., Giraldo, L.~O., Smith, G., Wolniewicz, K., Ackley, K., Salwen, C., Triolo, J., Pinelli, D., and Luong, K., ``{Front-end ASIC for spectroscopic readout of virtual Frisch-grid CZT bar sensors},'' {\em Nuclear Instruments and Methods in Physics Research Section A: Accelerators, Spectrometers, Detectors and Associated Equipment}~{\bf 940},  1--11 (2019).

\bibitem{Shy_2023}
Shy, D., Woolf, R.~S., Sleator, C.~C., Wulf, E.~A., Johnson-Rambert, M., Kong, E., Davis, J.~M., Caligiure, T.~J., Grove, J.~E., and Phlips, B.~F., ``Development of a csi calorimeter for the compton-pair (compair) balloon-borne gamma-ray telescope,'' {\em IEEE Transactions on Nuclear Science}~{\bf 70},  2329–2336 (Oct. 2023).

\bibitem{onsemiJ}
``Arrayj-series sipm sensors rev. 7..'' https://www.onsemi.com/pdf/datasheet/arrayj-series-d.pdf (September 2021).

\bibitem{rosspad}
``Rosspad.'' https://ideas.no/products/rosspad/ (2022).

\bibitem{onsemiC}
``C-series sipm sensors rev. 9..'' https://www.onsemi.com/pdf/datasheet/microc-series-d.pdf (February 2022).

\bibitem{MakotoTM}
Sasaki, M., ``{Trigger system for the ComPair instrument},'' in [{\em {Space Telescopes and Instrumentation 2020: Ultraviolet to Gamma Ray}}{\nolinebreak\hspace{0.1em}]},  den Herder, J.-W.~A., Nikzad, S., and Nakazawa, K., eds.,  {\bf 11444},  1023 -- 1028, International Society for Optics and Photonics, SPIE (2020).

\bibitem{switch}
``{LMX-1802G-SFP}.'' https://www.antaira.com/products/managed-gigabit/LMX-1802G-SFP.

\bibitem{mcp}
``Mcp9808.'' https://www.microchip.com/en-us/product/mcp9808 (2020).

\bibitem{megalib}
Zoglauer, A., Andritschke, R., and Schopper, F., ``Megalib – the medium energy gamma-ray astronomy library,'' {\em New Astronomy Reviews}~{\bf 50}(7),  629--632 (2006).
\newblock Astronomy with Radioactivities. V.

\bibitem{higs}
Litvinenko, V.~N., Burnham, B., Emamian, M., Hower, N., Madey, J. M.~J., Morcombe, P., O'Shea, P.~G., Park, S.~H., Sachtschale, R., Straub, K.~D., Swift, G., Wang, P., Wu, Y., Canon, R.~S., Howell, C.~R., Roberson, N.~R., Schreiber, E.~C., Spraker, M., Tornow, W., Weller, H.~R., Pinayev, I.~V., Gavrilov, N.~G., Fedotov, M.~G., Kulipanov, G.~N., Kurkin, G.~Y., Mikhailov, S.~F., Popik, V.~M., Skrinsky, A.~N., Vinokurov, N.~A., Norum, B.~E., Lumpkin, A., and Yang, B., ``Gamma-ray production in a storage ring free-electron laser,'' {\em Phys. Rev. Lett.}~{\bf 78},  4569--4572 (Jun 1997).

\bibitem{grape}
{O{\~n}ate Melecio}, K., {Bancroft}, C., {Ertley}, C., {Kislat}, F., {Legere}, J., {Longworth}, S., {McConnell}, M.~L., {Mello}, K., {Puopolo}, D., {Zaid}, J., and {Ryan}, J., ``{An advanced design for the Gamma-RAy Polarimeter Experiment (GRAPE)},'' in [{\em Hard X-Ray, Gamma-Ray, and Neutron Detector Physics XXIV}{\nolinebreak\hspace{0.1em}]},  {\em Society of Photo-Optical Instrumentation Engineers (SPIE) Conference Series} {\bf 12241},  122410G (Oct. 2022).

\bibitem{REGENER1935}
Regener, E. and Pfotzer, G., ``Vertical intensity of cosmic rays by threefold coincidences in the stratosphere,'' {\em Nature}~{\bf 136},  718--719 (Nov 1935).

\bibitem{kirschy}
Kirschner, N. et~al., ``{The Double-sided Silicon Strip Detector Tracker onboard the ComPair Balloon Flight },'' Space Telescopes and Instrumentation 2024: Ultraviolet to Gamma Ray, Paper 13093-297 (2024).

\bibitem{metzler}
Metzler, Z. et~al., ``{Balloon Flight Results and Calibration of the Anti-Coincidence Detector Subsystem for ComPair},'' Space Telescopes and Instrumentation 2024: Ultraviolet to Gamma Ray, Paper 13093-294 (2024).

\bibitem{shy24}
Shy, D. et~al., ``{Results from the CsI Calorimeter onboard the ComPair Balloon Flight},'' Space Telescopes and Instrumentation 2024: Ultraviolet to Gamma Ray, Paper 13093-301 (2024).

\bibitem{compair2}
Caputo, R. et~al., ``{ComPair 2: a Next Generation Medium Energy Gamma-ray Telescope Prototype},'' Space Telescopes and Instrumentation 2024: Ultraviolet to Gamma Ray, Paper 13093-94 (2024).

\bibitem{AMEGOX}
Caputo, R. et~al., ``{All-sky Medium Energy Gamma-ray Observatory eXplorer mission concept},'' {\em Journal of Astronomical Telescopes, Instruments, and Systems}~{\bf 8}(4),  044003 (2022).

\end{thebibliography}
\bibliographystyle{spiebib} 

\end{document}